# The Effect of Defect Layer on Lasing in Cholesteric Liquid Crystal

A.H. Gevorgyan[1,2], K. B. Oganesyan[3,4], N.Sh. Izmailian[3,5], E.A. Ayryan[4],
D.B. Mukhina[6], A.V. Zavozina[6], B. Lesbayev[7], G. Lavrelashvili[8], M. Hnatic[9,10],
Yu.V. Rostovtsev[11], G. Kurizki[12], M.O. Scully[13]

[1]Yerevan State University, 1 Al. Manookian St., 025, Yerevan, Armenia
[2]Ins. of Applied Problems in Physics, 26, Hr. Nersessian, 0014, Yerevan, Armenia
[3]Alikhanyan National Science Lab, Yerevan Physics Institute, Alikhanyan Br.2, 036, Yerevan, Armenia, bsk@.yerphi.am
[4]Laboratory of Information Technologies, JINR, Dubna, Russia
[5]Institute of Physics, Academia Sinica, Taipei, Taiwan
[6]Peopel's Friendship University of Russia, Moscow, Russia
[7]AInstitute of Combustion Problems, Almaty, Kazakhstan
[8]A.Razmadze Mathematical Institute, Tbilisi, Georgia
[9] Faculty of Sciences, P. J. Safarik University, Kosice, Slovakia
[10] Institute of Experimental Physics SAS, Kosice, Slovakia
[11]University of North Texas, Denton, TX, USA
[12]Weizmann Institute of Science, Rehotot, Israel
[13]Texas A&M University, College Station, Texas, USA

The photonic density of states (PDS) of eigen polarizations (EPs) in cholesteric liquid crystal (CLC) cells with a defect layer inside are calculated. The dependences for the PDS and light intensity in the defect layer on the parameters characterizing absorption and gain are obtained. We investigated the possibility of connections between the PDS and the density of the light energy accumulated in the system. The influence of the defect layer and CLC layer on the PDS are investigated. It is shown that the PDS is maximum when the defect is in the centre of the system. We showed also that the subject system can work as a low threshold laser, a multi-position trigger, filter, etc.

## 1. INTRODUCTION

Photonic crystals (PCs) consisting of artificial or self-organizing periodic structures have opened possibilities for novel physics and device applications. PCs as laser cavities have been the focus of intense research in recent decades. Lasing in these structures takes place either due to the distributed feedback mechanism or defect modes (in PCs with defects), where spontaneous emission is inhibited in the wavelengths inside the photonic band gaps (PBGs) except at the band edges or defect modes where the photonic density of states (PDS) is large. Especially attractive is the use of cholesteric liquid crystals (CLCs) as band edge lasers which is connected with the fact that the macroscopic properties of CLCs can be manipulated by external stimuli. Lasing in

CLC and PCs has been intensely investigated (see, for instance, [1-4] and the wide literature cited therein). The possibilities of decreasing the lasing threshold in CLC and multilayer systems with CLC layers have been studied both theoretically [5-7] and experimentally [8]. In the following three papers [9-11] it was reported on the effect of loss and gain on the PDS. Dolganov reported in [12] that a highly sensitive method was used to determine the PDS, as well as a drastic increase in the PDS near the PBG borders and oscillations related to Pendellösung beatings were observed. It is reported in [13] about design and fabrication of a wedge structured CLC film incorporating a spatial gradient of a chiral dopant concentration. A continuous spatial laser tuning in a broad visible spectral range with tuning resolution less than 1 nm is demonstrated, which renders a CLC-based micron-sized laser an important continuously tunable laser device [13]. In review article [14] the ways of employing CLCs in tunable dye lasers were discussed. The investigations in this direction have been intensely going on because the CLC, all the time, exhibit new wonderful and multilateral peculiarities (see, for example, [15-24]). The performance of light emitting liquid crystal devices may be significantly improved by using more advanced optical structures. To develop new optical architectures one must have at his disposal fundamental understanding of absorption-emission processing and photonic density of states and their other important optical peculiarities, as well as an accurate numerical design tools for their calculation.

In this paper we discussed lasing peculiarities of a CLC layer with an isotropic defect layer inside.

## 2. RESULTS AND DISCUSSION

A CLC with an isotropic defect can be treated as a three-layer system: two CLC layers (CLC(1) and CLC(2)) and an isotropic layer (IL) between them (or a Fabry–Perot resonator with diffraction mirrors and an isotropic layer filling). The ordinary and extraordinary refractive indices of the CLC layers are taken to be $n_o = \sqrt{\varepsilon_2} = 1.4639$ and $n_e = \sqrt{\varepsilon_1} = 1.5133$; $\varepsilon_1$, $\varepsilon_2$ are the principal values of the CLC local dielectric tensor. The CLC layer helix is right handed and its pitch is: $p = 420$ nm. These are the parameters of the CLC cholesteryl-nonanoate–cholesteryl-chloride–cholesteryl-acetate (20:15:6) composition at the temperature $t = 25^{\circ}$C. Hence, the light normally incident onto a single CLC layer – with the right circular polarization (RCP) – has a PBG (which is in the range of $\lambda = 614.8 \div 635.6$ nm), and the light with the left circular polarization (LCP) does not have any. The defect isotropic layer refractive index was taken to be $n =1.8$. The problem was solved by Ambartsumian's layer addition modified method [25]. We investigated the peculiarities of absorption, emission and PDS of this system for eigen polarizations (EPs). The EPs are the two polarizations of incident light, which do not change when light transmits through the system.

The presence of a thin defect in the CLC structure is known to initiate a defect mode in the PBG [26, 27]. First we investigate some new peculiarities of the PDS for this system. The investigation of the PDS $\rho$ is important because of the following. For laser emission it has been shown that, for instance, analyzing the case of the Fabry-Perot resonator, the threshold gain $g_{th}$ can be directly related to the maximum PDS $\rho_{max}$, that is, $g_{th} \propto n/\rho_{max}d$, where $n$ is the refractive index inside the resonator of the length $d$ [28]. Furthermore, according to the space-independent rate equations, the slope efficiency of lasers can be shown to be inversely proportional to the threshold energy and, therefore, directly proportional to $\rho_{max}$ [29]. The PDS, i.e. the number of wave vectors $k$ per unit frequency – $\rho(\omega) = dk/d\omega$ – is the reverse of the group velocity and can be defined by the expression [29]:

$$\rho_i(\omega) \equiv \frac{dk_i}{d\omega} = \frac{1}{d} \frac{\frac{du_i}{d\omega}v_i - \frac{dv_i}{d\omega}u_i}{u_i^2 + v_i^2}, \quad i=1,2 \tag{1}$$

where $d$ is the whole system thickness, $\omega$ is the incident light frequency, and $u_i$ and $v_i$ are the real and imaginary parts of the transmission amplitudes; $t_i(\omega) = u_i(\omega) + jv_i(\omega)$ are the transmission amplitudes for the incident light with the two EPs, $j$ is the imaginary unit. For the single CLC layer the two EPs practically coincide with the two circular (right and left) polarizations. The values $i=1, 2$ correspond to the diffracting and non-diffracting EPs, respectively. For a CLC layer with a defect layer inside a jump of eigen polarizations takes place at the defect mode. For the isotropic case we have: $\rho_{iso} = n_s/c$, where $n_s = \sqrt{\varepsilon_s}$ is the refractive index of the media surrounding the system and $c$ is the speed of light in vacuum.

One must realize that the concept of the PDS introduced by Eq. (1) is not without controversy, especially in the case of non-periodic systems (more details about this see in [30]). Let us only note that although it is impossible to ascribe a direct physical meaning to the PDS in Eq. (1) in general case, anyhow, it can be used as a parameter, which can provide some heuristic guidance in experiments for the PDS dispersion-related effects. Moreover, as it had been shown in [9-12] Eq. (1) is applicable for finite PCs layers, even for the cases with absorption and amplification (naturally, for weak ones).

Below we first show that this equation is applicable (at least qualitatively) to our (aperiodic) system both in the presence of absorption and amplification.

In [10-11, 30,31], it was shown that in the absence of absorption, the PDS was proportional to the integral of the energy stored within the medium. Then in [10, 11], it was shown that it takes place both in the presence of absorption and gain ( for more details see [33-100] ).

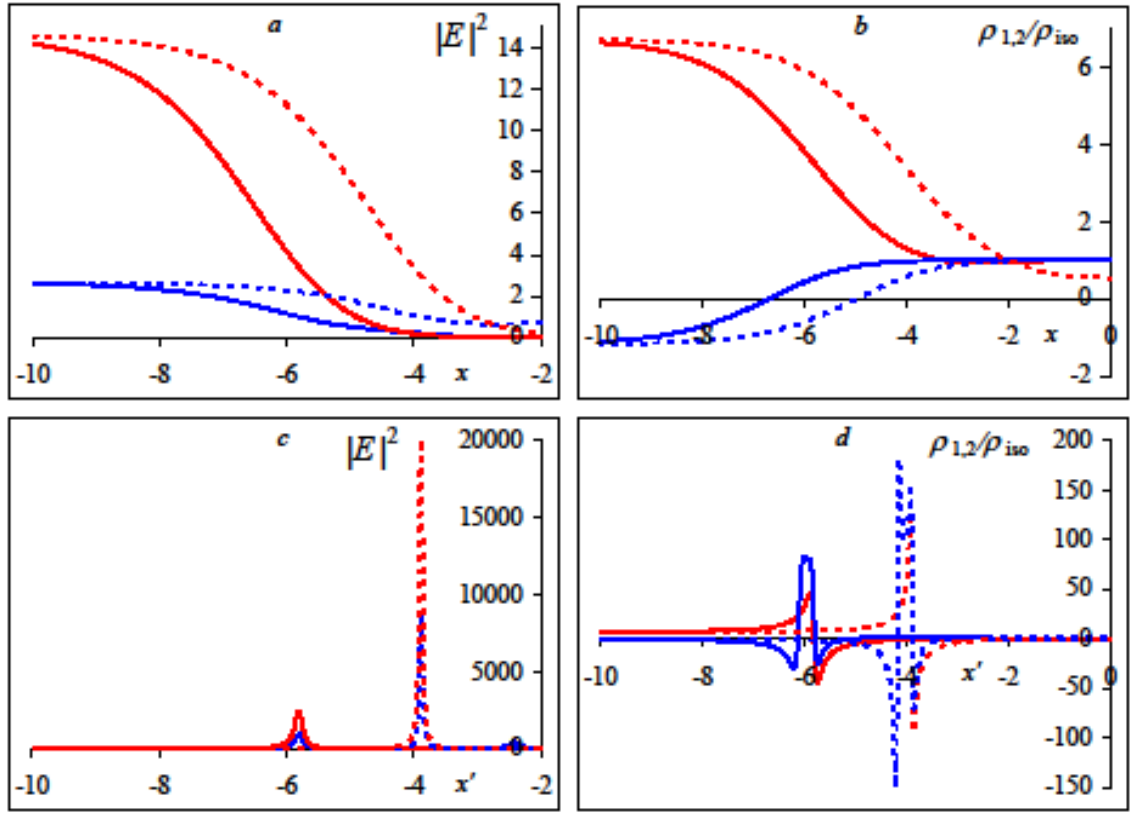

Fig. 1. The dependences (*a, c*) of the total field intensity $|E|^2$ aroused at the left border of the defect layer and (*b, d*) the relative PDS $\rho_{1,2}/\rho_{iso}$ (*a, b*) on the absorption parameter $x = \ln(\text{Im}\,\varepsilon_m)$ and (*c, d*) on the parameter $x' = \ln(-\text{Im}\,\varepsilon_m)$ characterizing gain at the given defect mode wavelength ($\lambda = 631.5 nm$) for the diffracting (the red lines) and non-diffracting (the blue lines) EPs. The dashed lines correspond to $n_o^" = n_e^" = 0$ and $n^" \neq 0$, that is, when absorption is absent in the CLC sublayers and it exists in the defect layer; and the solid lines correspond to the case when $n_o^" = n_e^" \neq 0$ and $n^` = 0$, that is, if absorption exists in the CLC sublayers and is absent in the defect layer. The CLC layer thickness is: $d=44p$, the defect layer thickness is: $d´=1000$nm.

In Fig. 1 the dependences (*a, c*) of the total field intensity $|E|^2$ aroused at left border of the defect layer and (*b, d*) the relative PDS $\rho_{1,2}/\rho_{iso}$ (*a, b*) on the absorption parameter $x = \ln(\text{Im}\,\varepsilon_m)$ and (*c ,d*) on the parameter $x' = \ln(-\text{Im}\,\varepsilon_m)$ characterizing the gain at the given defect mode wavelength $\lambda = 631.5 nm$ are presented for the diffracting (the red lines) and non-diffracting (the blue lines) EPs ($\varepsilon_m = \frac{\varepsilon_1 + \varepsilon_2}{2}$ is the mean dielectric permittivity of CLC). The dashed lines correspond to the case if $n_o^{''} = n_e^{''} = 0$ and $n^{''} \neq 0$, that is, if absorption is absent in the CLC sublayers and it is present in the defect layer; and the solid lines correspond to the case if $n_o^{''} = n_e^{''} \neq 0$ and $n^{''} = 0$, that is, if absorption is present in the CLC sublayers and is absent in the defect layer.

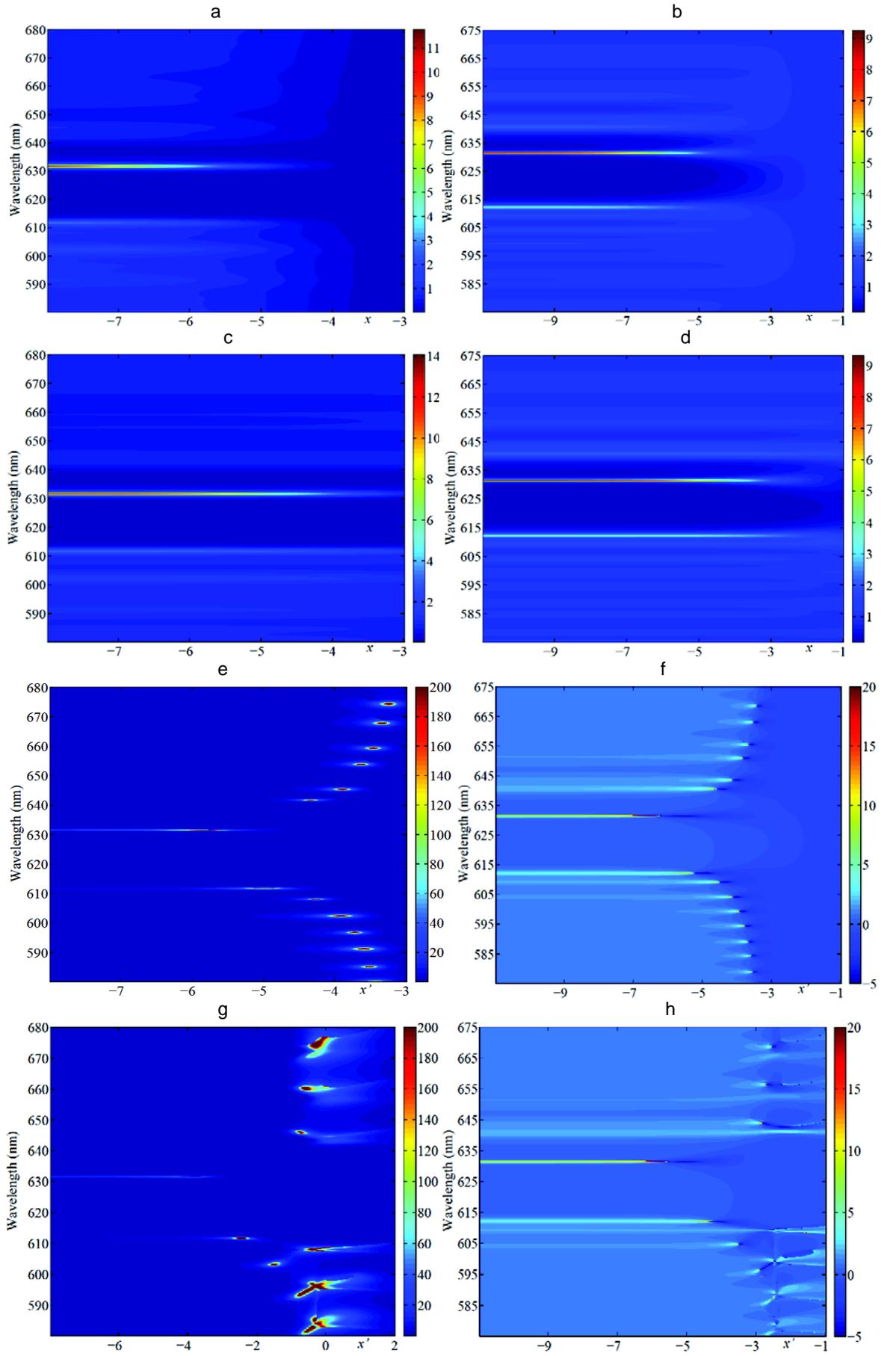

Fig.2. The evolution (*a, c*) of the total field intensity $|E|^2$ spectra and (*b, d*) the PDS $\rho_1 / \rho_{iso}$ spectra if (*a, b, c, d*) the absorption parameter $x = \ln(\text{Im}\,\varepsilon_m)$ increases and if (*e, f, g, h*) the gain parameter $x' = \ln(-\text{Im}\,\varepsilon_m)$

increases, for the diffracting EP. (*a, b, e, f*) correspond to case $n_o^{''} = n_e^{''} = 0$ and $n^{''} \neq 0$, and (*c, d, g, h*) correspond to the case if $n_o^{''} = n_e^{''} \neq 0$ and $n^{''} = 0$. The parameters are the same as in Fig.1.

In Fig.2 the evolution (*a, c*) of the total field intensity $|E|^2$ spectra and (*b, d*) the relative PDS $\rho_1 / \rho_{iso}$ spectra when (*a, b, c, d*) the absorption parameter $x = \ln(\operatorname{Im} \varepsilon_m)$ increases and when (*e, f, g, h*) the gain parameter $x' = \ln(-\operatorname{Im} \varepsilon_m)$ increases are presented for the diffracting EP. (*a, b, e, f*) correspond to case if $n_o^{''} = n_e^{''} = 0$ and $n^{''} \neq 0$, and (*c, d, g, h*) correspond to the case if $n_o^{''} = n_e^{''} \neq 0$ and $n^{''} = 0$.

The presented results show that at sharp changes of $|E|^2$ and, consequently, also for the energy stored within the system, depending on the parameters *x* and *x´*, the expression $\rho_1 / \rho_{iso}$ also possesses sharp changes versus the same parameters. This means that Eq. (1) can also describe laser peculiarities of periodic systems with a defect in their structure. At the same time this figures give information about laser peculiarities of CLC layers with defects inside them in the presence of absorption and amplification. In particular, as can be seen in Fig. 1, the modulus of the PDS is decreasing dramatically with increase of the absorption parameter $x = \ln(\operatorname{Im} \varepsilon_m)$, as it takes place for a single CLC layer at the band gap border wavelengths. The maximum of the PDS increases with the gain increase. The further increase of gain leads to a resonance-like change with the maximum spikes in the PDS. There exists a critical value of *x'* beyond which the lasing mode is quenched and the feedback vanishes. Note that the critical values of *x'* for the case if gain is present in the CLC sublayers and that for the case if it is absent in the defect layer – and vice versa – are different. Existence of these critical values of gain indicates the possibility that the subject system can work as a trigger. Tuning the pumping intensity by a suitable dye molecule density or by the pump laser intensity, one can pass from the laser generation regime to the quenching one.

For a single CLC layer, as gain increases in the absence of the defect layer, the maximum PDS shifts away from the PBG border, and this does not take place continuously, but in discrete steps [10, 11]. Note that when going away from the PBG borders, the critical value of *x'* (beyond which the lasing mode is quenched) increases. As it is seen in Fig.2*e*, if a defect layer is present and gain increases, the PDS increases too, and then quenching takes place. But first it takes place for the defect mode wavelength. It is to be noted that not all modes outside the PBG exist in the case if gain is absent in the CLC layers and it exists in the defect layer. Again, let us note that such changes take place for $|E|^2$ too.

Then we considered the influence of the defect layer thickness on the relative PDS. In Fig.3 the evolution of the relative PDS – $\rho_{1,2}/\rho_{iso}$ – spectra when the defect layer thickness increases are presented for (*a*) the diffracting and (*b*) non-diffracting EPs. As is well known the defect mode wavelength increases from a minimum to a maximum band gap value if the defect layer optical thickness increases. As we can see in this figure, for this change of the defect layer thickness, the relative PDS $\rho_{1,2}/\rho_{iso}$ on the defect mode does not change smoothly, but with strong oscillations. And this means that there exists a definite value of the defect layer thickness at which the lowest threshold lasing takes place.

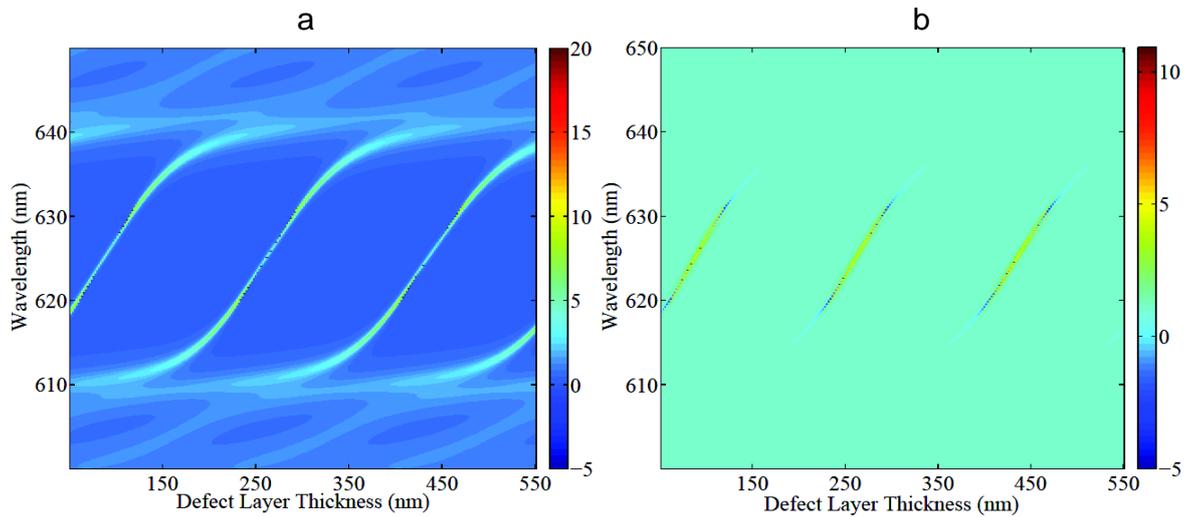

Fig.3. The density plots for the PDS spectra versus the defect layer thickness. $d'=3000$nm. The other parameters are the same as in Fig.1.

Now we pass to investigate the influence of the CLC layer thickness on the relative PDS. In Fig.4 the evolution of the relative PDS spectra if the CLC layer thickness increases are presented for the (a) diffracting and (b) non-diffracting EPs. As is seen in the figure, though low threshold lasing takes place both for diffracting and non-diffracting EPs only for small CLC layer thicknesses, nevertheless, for CLC layer large thicknesses it takes place only for the non-diffracting EP.

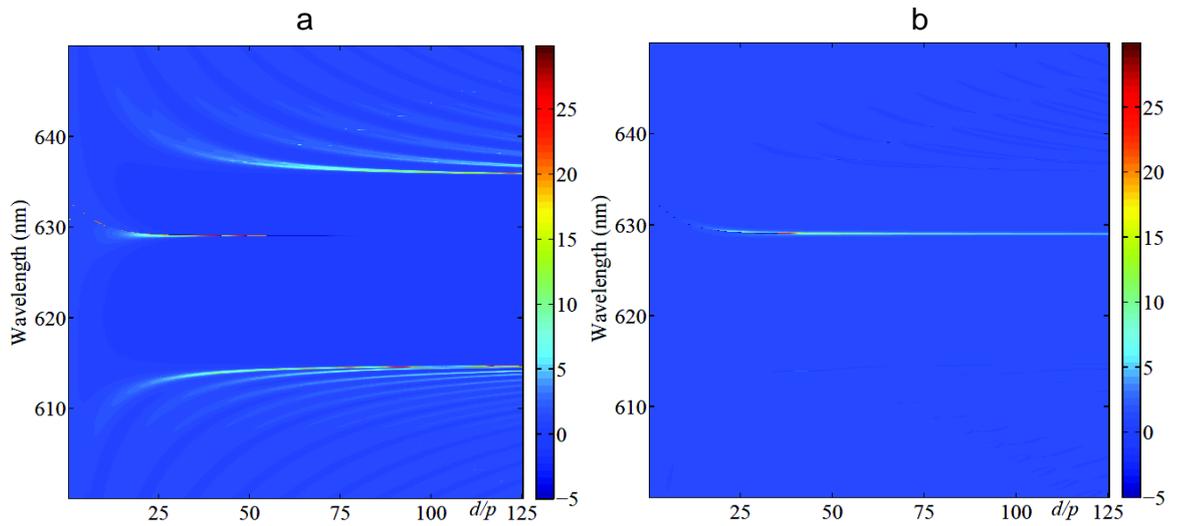

Fig.4. The density plots for the PDS spectra versus the CLC layer thickness. $d´=2380$nm. The other parameters are the same as in Fig.1.

Let us now consider how the position of the defect layer in the system affects on the defect mode features. Figure 5 shows the dependences of the relative PDS on the defect position in the structure ($z/p$, where $z$ is counted from the left boundary the CLC layer) for the (a) first and (b) second EPs. As is seen in the figure, laser generation with the least threshold takes place if the defect layer is in the CLC layer center. And it is seen in the figure that the defect location significantly influences on the PBG width too, essentially influencing on the PBG border near where the defect mode is.

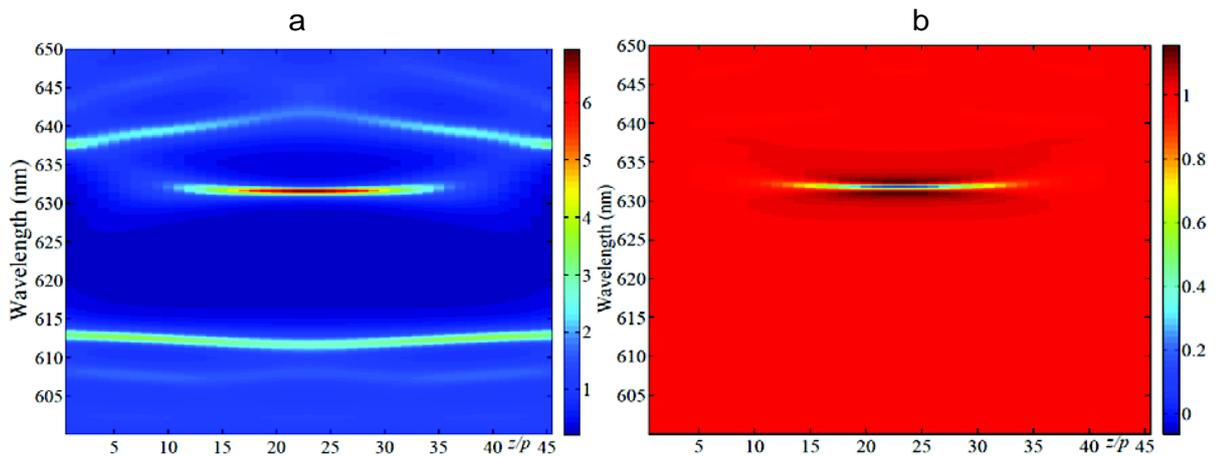

Fig.5. The density plots for the PDS spectra versus the defect layer position ($z/p$). The parameters are the same as in Fig.1.

## 3. CONCLUSIONS

Let us note in conclusion that we investigated lasing peculiarities of a CLC layer with a defect inside its structure. We showed that the PDS defined in [29] are applicable (qualitatively, at least) to the subject system, too. In the case of the CLC layer with an isotropic defect layer inside on the defect mode, the $\rho_1/\rho_{iso}$ has extreme – at the defect mode wavelength – with a larger height than that at the PBG borders. This provides laser generation with a much lower threshold at the defect modes than that at the PBG borders. Then we investigated the influence of absorption and gain on the PDS and the light field intensity in the defect. Our investigation of the influence of the defect layer thickness on the PDS showed that at the defect layer thickness changes the relative PDS $\rho_{1,2}/\rho_{iso}$ does not change smoothly, but with strong oscillations. Furthermore, when investigating the influence of the CLC layer thickness on the PDS we showed that though for small CLC layer thicknesses low threshold lasing takes place both for diffracting and non-diffracting EPs, nevertheless, for larger thicknesses it takes place only for the non-diffracting EP. Finally, we investigated the defect location influence on lasing peculiarities of the system.